\documentclass[letterpaper, 10 pt, conference]{ieeeconf}
\IEEEoverridecommandlockouts                              % This command is only needed if
\usepackage{amsmath,amsfonts,amssymb}%
\usepackage{bm}%
\usepackage{graphicx}%
\usepackage{cases}%
\usepackage{cite,url,citesort}
\usepackage{verbatim,latexsym}

\newtheorem{definition}{Definition}

\newtheorem{theorem}{Theorem}
\newtheorem{remark}{Remark}

\newtheorem{lemma}{Lemma}

\begin{document}
%
% paper title
% can use linebreaks \\ within to get better formatting as desired
\title{\LARGE \bf
 Performance Analysis and Coherent Guaranteed Cost Control for Uncertain Quantum Systems Using Small Gain and Popov Methods}

\author{Chengdi Xiang, Ian R. Petersen and Daoyi Dong
\thanks{Chengdi Xiang, Ian R. Petersen and Daoyi Dong are with the School of Engineering and Information Technology, University of New South Wales at the Australian Defence Force Academy, Canberra ACT 2600, Australia. {\tt\small \{elyssaxiang, i.r.petersen, daoyidong\}@gmail.com} This work was supported by the Australian Research Council.}%
}

%\author{\IEEEauthorblockN{Michael Shell}
%\IEEEauthorblockA{School of Electrical and\\Computer Engineering\\
%Georgia Institute of Technology\\
%Atlanta, Georgia 30332--0250\\
%Email: http://www.michaelshell.org/contact.html}
%\and
%\IEEEauthorblockN{Homer Simpson}
%\IEEEauthorblockA{Twentieth Century Fox\\
%Springfield, USA\\
%Email: homer@thesimpsons.com}

\maketitle

\begin{abstract}
%\boldmath
This paper extends applications of the quantum small gain and Popov methods from existing results on robust stability to performance analysis results for a class of uncertain quantum systems. This class of systems involves a nominal linear quantum system and is subject to quadratic perturbations in the system Hamiltonian. Based on these two methods, coherent guaranteed cost controllers are designed for a given quantum system to achieve improved control performance. An illustrative example also shows that the quantum Popov approach can obtain less conservative results than the quantum small gain approach for the same uncertain quantum system.
\end{abstract}

\IEEEpeerreviewmaketitle

\section{INTRODUCTION}\label{introduction}

%\hfill January 11, 2007

%-------------------
Due to recent advances in quantum and nano-technology, there has been a considerable attention focusing on research in the area of quantum  feedback control systems; e.g., [1]-[24]. In particular, robust control has been recognized as a critical issue in quantum control systems, since many quantum systems are unavoidably subject to disturbances and uncertainties in practical applications; e.g.,\cite{matt2008}, \cite{Ian2010} and \cite{Dong2010}. A majority of papers in the area of quantum feedback control only consider the case in which the controller is a classical system. In this case, analog and digital electronic devices may be involved and quantum measurements are required in the feedback loop. Due to the limitations imposed by quantum mechanics on the use of quantum measurement, recent research has considered the design of coherent quantum controllers to achieve improved performance. In this case, the controller itself is a quantum system; e.g., \cite{matt2008}, \cite{M2003} and \cite{2M2003}. In the linear case, the quantum system is often described by linear quantum stochastic differential equations (QSDEs) that require physical realizability conditions in terms of constraints on the system matrices to represent physically meaningful systems.

As opposed to using QSDEs, the papers \cite{matt2010},\cite{matt2009} have introduced a set of parameterizations $(S,L,H)$ to represent a class of open quantum systems, where $S$ is a scattering matrix, $L$ is a vector of coupling operators and $H$ is a Hamiltonian operator. The matrix $S$, together with the vector $L$, describes the interface between the system and the field, and the operator $H$ represents the energy of the system. The advantage of using a triple $(S,L,H)$ is that this framework automatically represents  a physically realizable quantum system. Therefore, in this paper, a coherent guaranteed cost controller is designed based on $(S,L,H)$ and the physical realizability condition does not need to be considered.

The small gain theorem and the Popov approach are two of the most important methods for the analysis of robust stability in classical control.
The paper \cite{Ian2012} has applied the small gain method to obtain a robust stability result for uncertain quantum systems. This result gives a sufficient condition for robust stability in terms of a strict bounded real condition. The small gain method has also been extended to the robust stability analysis of quantum systems with different perturbations and applications (e.g., see \cite{IanMaui}, \cite{IanDubrovnik}, \cite{IanTurkey} and \cite{IanFlorence}). The paper \cite{Valery} has introduced a quantum version of the Popov stability criterion in terms of a frequency domain condition which is of the same form as the classical Popov stability condition (e.g., see \cite{nonlinear}). The Popov approach has also been used to analyze the robust stability of an optical cavity containing a saturated Kerr medium \cite{IanZurich}. Also, the paper \cite{Valery} has shown that the frequency domain condition obtained in \cite{Valery} is less conservative than the stability result using the small gain theorem \cite{Ian2012}.

In this paper, we extend the quantum small gain method in \cite{Ian2012} and the Popov type approach in \cite{Valery} from robust stability analysis to robust performance analysis for uncertain quantum systems. We assume that the system Hamiltonian $H$ can be decomposed in terms of a known nominal Hamiltonian $H_1$ and a perturbation Hamiltonian $H_2$, i.e., $H=H_1+H_2$. The perturbation Hamiltonian $H_2$ is contained in a set of Hamiltonians $\mathcal{W}$. We consider uncertain quantum systems where the nominal system is a linear system and the perturbation Hamiltonian is quadratic. Moreover, a coherent controller is designed using the small gain approach and the Popov approach for the uncertain quantum system, where a guaranteed bound on a cost function is derived in terms of linear matrix inequality (LMI) conditions. Although preliminary versions of the results in this paper have been presented in the conference papers \cite{Chengdi1} and \cite{Chengdi2}, this paper presents complete proofs of the main results and modifies the example in \cite{Chengdi1} for a consistent performance comparison of the proposed two methods.

The remainder of the paper proceeds as follows. In Section \ref{quantum systems}, we define the general class of quantum systems under consideration and specify the nominal system as a linear quantum system. We then present a set of quadratic Hamiltonian perturbations in Section \ref{perturbations of the hamiltonian}. In Section \ref{performance analysis}, a performance cost function for the given system is defined. Moreover, a small gain approach and a Popov type method are used to analyze the performance of the given system. In Section \ref{controller design}, a quantum controller is added to the original system to stabilize the system and also to achieve improved performance. Also, corresponding approaches are used to construct a coherent guaranteed cost controller in terms of LMI conditions. In Section \ref{illustrative example}, an illustrative example is provided to demonstrate the method that is developed in this paper. A performance comparison between these two methods is also shown in the illustrative example. We present some conclusions in Section \ref{conclusion}.

\section{QUANTUM SYSTEMS}\label{quantum systems}
%%------------------------------------------------
In this section, we describe the general class of quantum systems under consideration, which is defined by parameters $(S,L,H)$. Here $H=H_1+H_2$, $H_1$ is a known self-adjoint operator on the underlying Hilbert space referred to as the nominal Hamiltonian and $H_2$ is a self-adjoint operator on the underlying Hilbert space referred to as the perturbation Hamiltonian contained in a specified set of Hamiltonians $\mathcal{W}$; e.g.,\cite{matt2010}, \cite{matt2009}. The set $\mathcal{W}$ can correspond to a set of exosystems (see, \cite{matt2010}). The corresponding generator for this class of quantum systems is given by
\begin{equation}
\mathcal{G}(X)=-i[X,H]+\mathcal{L}(X)
\end{equation}
where $\mathcal{L}(X)=\frac{1}{2}L^\dag[X,L]+\frac{1}{2}[L^\dag,X]L$.
Here, $[X,H]=XH-HX$ describes the commutator between two operators and the notation $^\dag $ refers to the adjoint transpose of a vector of operators. Based on a quantum stochastic differential equation, the triple $(S,L,H)$, together with the corresponding generators, defines the Heisenberg evolution $X(t)$ of an operator $X$ \cite{matt2010}. The results presented in this paper will build on the following results from \cite{Valery}.
%above sentence need some rephrase
%-------------------------------
\begin{lemma}\label{baselemma}
\cite{Valery} Consider an open quantum system defined by $(S, L, H)$ and suppose there exist non-negative self-adjoint operators $V$ and $W$ on the underlying Hilbert space such that
\begin{equation}\label{lemma1}
\mathcal {G}(V)+W\leq \lambda
\end{equation}
where $\lambda$ is a real number. Then for any plant state, we have
\begin{equation}
\limsup \limits_{T\to \infty} \frac{1}{T}\int_0^T\langle W(t)\rangle dt\leq\lambda.
\end{equation}
Here $W(t)$ denotes the Heisenberg evolution of the operator $W$ and $\langle\cdot\rangle$ denotes quantum expectation; e.g., see \cite{Valery} and \cite{matt2010}.
\end{lemma}

%%------------------------------------------------

In this paper, the nominal system is considered to be a linear quantum system. We assume that $H_1$ is in the following form
\begin{equation}\label{17}
H_1=\frac{1}{2}\left[\begin{array}{c c} a^\dag & a^T\end{array}\right]M\left[\begin{array}{l} a \\ a^\#\end{array}\right]
\end{equation}
where $M\in \mathbb{C}^{2n\times2n}$  is a Hermitian matrix and has the following form with $M_1=M^\dag_1$ and $M_2=M^T_2$
\begin{equation}
M=\left[\begin{array}{cc}M_1&M_2\\M_2^\#&M_1^\#\end{array}\right].
\end{equation}
Here $a$ is a vector of annihilation operators on the underlying Hilbert space and $a^\#$ is the corresponding vector of creation operators. In the case of matrices, the notation $ ^\dag$ refers to the complex conjugate transpose of a matrix. In the case of vectors of operators, the notation $ ^\#$ refers to the vector of adjoint operators and in the case of complex matrices, this notation refers to the complex conjugate matrix.
The canonical commutation relations between annihilation and creation operators are described in the following way
\begin{equation}\label{commutation1}
\begin{split}
\left[\begin{array}{c}\left[\begin{array}{l}a\\a^\#\end{array}\right] ,\left[\begin{array}{l}a\\a^\#\end{array}\right]^\dag \end{array}\right]=&\left[\begin{array}{l}a\\a^\#\end{array}\right] \left[\begin{array}{l}a\\a^\#\end{array}\right]^\dag\\
&-\left(\begin{array}{c}\left[\begin{array}{l}a\\a^\#\end{array}\right]^\#\left[\begin{array}{l}a\\a^\#\end{array}\right]^T \end{array}\right)^T\\
=&\ J
\end{split}
\end{equation}
where $J=\left[\begin{array}{cr}I&0\\0&-I \end{array} \right]$ \cite{Ian2010}.
%could add more reference for the above one.

The coupling vector $L$ is assumed to be of the form
\begin{equation}\label{19}
L=\left[\begin{array}{c c} N_1&N_2\end{array}\right]\left[\begin{array}{l} a \\ a^\#\end{array}\right]
=\tilde{N}\left[\begin{array}{l} a \\ a^\#\end{array}\right]
\end{equation}
where $N_1\in \mathbb{C}^{m\times n}$ and $N_2\in \mathbb{C}^{m\times n}$.
We also write
\begin{equation}
\left[\begin{array}{l}L\\L^\#\end{array}\right]=N\left[\begin{array}{l}a\\a^\#\end{array}\right]=\left[\begin{array}{c c} N_1&N_2\\N^\#_2&N^\#_1\end{array}\right]\left[\begin{array}{l} a \\ a^\#\end{array}\right].
\end{equation}
When the nominal Hamiltonian $H$ is a quadratic function of the creation and annihilation operators as shown in (\ref{17}) and the coupling operator vector is a linear function of the creation and annihilation operators, the nominal system corresponds to a linear quantum system (see, \cite{Ian2010}).

We consider self-adjoint ``Lyapunov" operators $V$ of the form
\begin{equation}\label{v}
V=\left[\begin{array}{c c} a^\dag & a^T\end{array}\right]P\left[\begin{array}{l} a \\ a^\#\end{array}\right]
\end{equation}
where $P\in \mathbb{C}^{2n\times2n}$  is a positive definite Hermitian matrix of the form
\begin{equation}\label{18}
P=\left[\begin{array}{cc}P_1&P_2\\P_2^\#&P_1^\#\end{array}\right].
\end{equation}

We then consider a set of non-negative self-adjoint operators $\mathcal{P}$ defined as
\begin{equation}
\mathcal{P}=\left\{\begin{array}{c} V\ \textnormal{ of the form} \ (\ref{v})\ \textnormal{such that} \ P>0\ \textnormal {is a}\\ \textnormal{Hermitian matrix of the form}\ (\ref{18})\end{array} \right\}.
\end{equation}
%%------------------------------------------------

%--------------------------------------------------------------------------------------------------------------

\section{PERTURBATIONS OF THE HAMILTONIAN}\label{perturbations of the hamiltonian}
In Section \ref{quantum systems}, we introduced the nominal linear quantum system. This section defines the quadratic Hamiltonian perturbations (e.g., see \cite{Ian2012}, \cite{Ian2012journal}) for the quantum system under consideration. We first define two general sets of Hamiltonians in terms of a  commutator decomposition, and then present two specific sets of quadratic Hamiltonian perturbations.
%-----------------------------------------------------------------------------------------
\subsection{Commutator Decomposition}
For the set of non-negative self-adjoint operators $\mathcal{P}$ and given real parameters $\gamma>0, \delta\geq0$, a particular set of perturbation Hamiltonians $\mathcal{W}_1$ is defined in terms of the  commutator decomposition
\begin {equation}\label{decomposition11}
[V,H_2]=[V, z^\dag]w-w^\dag[z,V]
\end{equation}
for $V\in \mathcal{P}$, where $w$ and $z$ are given vectors of operators.
$\mathcal{W}_1$ is then defined in terms of sector bound condition:
\begin{equation}\label{sector11}
w^\dag w\leq\frac{1}{\gamma^2}z^\dag z+\delta.
\end{equation}
We define
%\begin{equation}
%\begin{split}
%\mathcal{W}_1=H_2, \exists w, z \textnormal {such that (\ref{sector1}) is satisfied} \\
%\textnormal{and (\ref{decomposition1})  is satisfied}  \forall V\in P.
%\end{split}
%\end{equation}
\begin{equation}
\mathcal{W}_1=\left\{\begin{array}{c}H_2 :\exists \ w, z\ \textnormal{such that} \ (\ref{sector11})\ \textnormal {is satisfied}\\ \textnormal{and}\ (\ref{decomposition11})\ \textnormal{is satisfied}\ \forall \ V\in \mathcal{P}\end{array} \right\}.
\end{equation}
%%%%%
\subsection{Alternative Commutator Decomposition}
Given a set of non-negative operators $\mathcal{P}$, a self-adjoint operator $H_1$, a coupling operator $L$, real parameters $\beta\geq0$ $\gamma>0$, and a set of Popov scaling parameters $\Theta\subset[0,\infty)$, we define a set of perturbation Hamiltonians $\mathcal{W}_2$ in terms of the commutator decompositions \cite{Valery}
\begin{equation}\label{sector1}
\begin{split}
[V-\theta H_1,H_2]&=[V-\theta H_1, z^\dag]w-w^\dag[z,V-\theta H_1],\\
\mathcal{L}(H_2)&\leq\mathcal{L}(z^\dag)w+w^\dag\mathcal{L}(z)+\beta[z,L]^\dag[z,L]\\
\end{split}
\end{equation}
for $V\in\mathcal{P}$ and $\theta\in\Theta$, where $w$ and $z$ are given vectors of operators. Note that (\ref{decomposition11}) and (\ref{sector1}) correspond to a general quadratic perturbation of the Hamiltonian.
This set $\mathcal{W}_2$ is then defined in terms of the sector bound condition
\begin{equation}\label{decomposition1}
(w-\frac{1}{\gamma}z)^\dag(w-\frac{1}{\gamma}z)\leq\frac{1}{{\gamma}^2}z^\dag z.
\end{equation}
We define
\begin{equation}\label{w1}
\mathcal{W}_2=\left\{\begin{array}{c}H_2\geq0 :\exists \ w, z\ \textnormal{such that} \ (\ref{sector1})\ \textnormal{and}\ (\ref{decomposition1})\\ \textnormal{are satisfied}\ \forall \ V\in \mathcal{P},\ \theta\in\Theta\end{array} \right\}.
\end{equation}
\subsection{Quadratic Hamiltonian Perturbation}
We consider a set of quadratic perturbation Hamiltonians that is in the form
\begin{equation}\label{h2form1}
H_2=\frac{1}{2}\left[\begin{array}{c c} \zeta^\dag & \zeta^T\end{array}\right]\Delta\left[\begin{array}{l} \zeta \\ \zeta^\#\end{array}\right]
\end{equation}
where $\zeta=E_1a+E_2a^\#$ and $\Delta\in\mathbb{C}^{2m\times2m}$ is a Hermitian matrix of the form
\begin{equation}\label{delta1}
\Delta=\left[\begin{array}{cc}\Delta_1&\Delta_2\\ \Delta^\#_2&\Delta^\#_1\end{array}\right]
\end{equation}
with $\Delta_1=\Delta^\dag_1$ and $\Delta_2=\Delta^T_2$.

Since the nominal system is linear, we use the relationship:
\begin{equation}\label{quadraticz}
z=\left[\begin{array}{l} \zeta \\ \zeta^\#\end{array}\right]=\left[\begin{array}{c c} E_1&E_2\\E^\#_2&E^\#_1\end{array}\right]\left[\begin{array}{l} a \\ a^\#\end{array}\right]=E\left[\begin{array}{l} a \\ a^\#\end{array}\right].
\end{equation}
Then
\begin{equation}
H_2=\frac{1}{2}\left[\begin{array}{c c} a^\dag & a^T\end{array}\right]E^\dag\Delta E\left[\begin{array}{l} a \\ a^\#\end{array}\right].
\end{equation}

When the matrix $\Delta$ is subject to the norm bound
\begin{equation}\label{delta_bound1}
\| \Delta \|\leq\frac{2}{\gamma},
\end{equation}
where $\|.\|$ refers to the matrix induced norm, we define
\begin{equation}
\mathcal{W}_3=\left\{\begin{array}{c}H_2 \textnormal { of the form (\ref{h2form1}) and (\ref{delta1}) such that }\\
\textnormal{condition (\ref{delta_bound1}) is satisfied}\end{array} \right\}.
\end{equation}
In \cite{Ian2012}, it has been proven that for any set of self-adjoint operators $\mathcal{P}$,
\begin{equation}\label{w1relation}
\mathcal{W}_3\subset\mathcal{W}_1.
\end{equation}
%-------------------------------
%%%%%%%%%%%%

%

When the matrix $\Delta$ is subject to the bounds
\begin{equation}\label{delta_bound}
0\leq\Delta\leq\frac{4}{\gamma}I,
\end{equation}
we define
\begin{equation}
\mathcal{W}_4=\left\{\begin{array}{c}H_2 \textnormal { of the form (\ref{h2form1}) and (\ref{delta1}) such that }\\
\textnormal{condition (\ref{delta_bound}) is satisfied}\end{array} \right\}.
\end{equation}
In \cite{Valery}, it has been proven that if $[z,L]$ is a constant vector, then for any set of self-adjoint operators $\mathcal{P}$,
\begin{equation}\label{w2relation}
\mathcal{W}_4\subset\mathcal{W}_2.
\end{equation}

%%-----------------------------------------------------------------------------------------------------------------------------

\section{PERFORMANCE ANALYSIS}\label{performance analysis}
%%------------------------------------------------
In this section, we provide several results on performance analysis for quantum systems subject to a quadratic perturbation Hamiltonian. Also, the associated cost function is defined in the following way:
\begin{equation}
\mathcal{J}=\limsup \limits_{T\to \infty} \frac{1}{T}\int_0^T\langle \left[\begin{array}{c c} a^\dag & a^T\end{array}\right]R\left[\begin{array}{l} a \\ a^\#\end{array}\right]\rangle dt
\end{equation}
where $R>0$.
We denote
\begin{equation}
W=\left[\begin{array}{c c} a^\dag & a^T\end{array}\right]R\left[\begin{array}{l} a \\ a^\#\end{array}\right].
\end{equation}
In order to prove the following  theorems on performance analysis, we require some algebraic identities.
%----------------------------------------------------------
\begin{lemma}\label{algebra}
(See Lemma 4 of\cite{Ian2012}) Suppose $V\in\mathcal{P}, H_1$ is of the form (\ref{17}) and $L$ is of the form (\ref{19}). Then
\begin{equation}
[V, H_1]= \left[\begin{array}{l}a\\a^\#\end{array}\right]^\dag(PJM-MJP) \left[\begin{array}{l}a\\a^\#\end{array}\right],
\end{equation}
\begin{align}
\mathcal{L}(V)=&-\frac{1}{2} \left[\begin{array}{l}a\\a^\#\end{array}\right]^\dag(N^\dag JNJP+PJN^\dag JN)\left[\begin{array}{l}a\\a^\#\end{array}\right]\nonumber\\
&+\textnormal{Tr}(PJN^\dag \left[\begin{array}{cc} I&0\\0&0\end{array}\right]NJ),
\end{align}
\begin{equation}
[\left[\begin{array}{l}a\\a^\#\end{array}\right],\left[\begin{array}{l}a\\a^\#\end{array}\right]^\dag P\left[\begin{array}{l}a\\a^\#\end{array}\right]]=2JP\left[\begin{array}{l}a\\a^\#\end{array}\right].
\end{equation}
\end{lemma}
%------------------------------------------------------------
\begin{lemma}\label{quadratic_algebra}
For $V\in\mathcal{P}$ and $z$ defined in (\ref{quadraticz}),
\begin{equation}
[z,V]=2EJP\left[\begin{array}{l} a\\a^\# \end{array}\right],
\end{equation}

\begin{equation}
[V,z^\dag][z,V]=4\left[\begin{array}{l} a\\a^\# \end{array}\right]^\dag PJE^\dag EJP\left[\begin{array}{l} a\\a^\# \end{array}\right],
\end{equation}

\begin{equation}
z^\dag z=\left[\begin{array}{l} a\\a^\# \end{array}\right]^\dag E^\dag E\left[\begin{array}{l} a\\a^\# \end{array}\right].
\end{equation}
\end{lemma}
\emph{Proof:}
The proof follows from Lemma \ref{algebra}.
\hfill $\Box$\\
%---------------------------------------------------------
\begin{lemma} (See Lemma 5 of \cite{Valery})\label{algebra2}
For $z$ defined in (\ref{quadraticz}) and $L$ being of the form (\ref{19}),
\begin{equation}
[z,L]=[E\left[\begin{array}{l}a\\a^\#\end{array}\right],\tilde{N}\left[\begin{array}{l}a\\a^\#\end{array}\right]]=EJ\Sigma\tilde{N}^T
\end{equation}
is a constant vector, where
\begin{equation}
\Sigma=\left[\begin{array}{ll}0&I\\I&0\end{array}\right].
\end{equation}
\end{lemma}
%----------------------------------------------------------------
\begin{lemma} (See Lemma 6 of \cite{Valery})\label{algebra3}
For $z$ defined in (\ref{quadraticz}), $H_1$ defined in (\ref{17}) and $L$ being of the form (\ref{19}),
we have
\begin{equation}
-i[z, H_1]+\mathcal{L}(z)=E(-iJM-\frac{1}{2}JN^\dag JN)\left[\begin{array}{l} a \\ a^\#\end{array}\right]
\end{equation}
and
\begin{equation}
i[z,V]=2iEJP\left[\begin{array}{l} a \\ a^\#\end{array}\right].
\end{equation}

\end{lemma}
%------------------------------------------------------------------
Now we present two theorems (Theorem 1 and Theorem 2) which can be used to analyze the performance of the given quantum systems using a quantum small gain method and a Popov type approach, respectively.
%--------------------------
\subsection{Performance Analysis Using the Small Gain Approach}
%explain performance
%In order to analyse the performance of the uncertain linear quantum systems with quadratic Hamiltonian perturbation, we need the following lemma.
\begin{theorem}\label{theorem1}
Consider an uncertain quantum system $(S, L, H)$, where $H=H_1+H_2$, $H_1$ is in the form of (\ref{17}),
$L$ is of the form (\ref{19}) and $H_2\in\mathcal{W}_3$. If $F=-iJM-\frac{1}{2}JN^\dag JN$ is Hurwitz,
\begin{equation}\label{riccati1}
\left[\begin{array}{cc}F^\dag P+PF+\frac{E^\dag E}{\gamma^2\tau^2}+R&2PJE^\dag\\ 2EJP&-I/\tau^2\end{array}\right]<0
\end{equation}
has a solution $P>0$ in the form of (\ref{18}) and $\tau>0$, then
\begin{align}
\mathcal{J}&=\limsup \limits_{T\to \infty} \frac{1}{T}\int_0^T\langle W(t)\rangle dt\nonumber\\
&=\limsup \limits_{T\to \infty} \frac{1}{T}\int_0^T\langle \left[\begin{array}{c c} a^\dag & a^T\end{array}\right]R\left[\begin{array}{l} a \\ a^\#\end{array}\right]\rangle dt\leq\tilde{\lambda}+\frac{\delta}{\tau^2}
\end{align}
where
\begin{equation}
\tilde{\lambda}=\text{Tr}(PJN^\dag \left[\begin{array}{cc} I&0\\0&0\end{array}\right]NJ).
\end{equation}
\end{theorem}
%%------------------------------------------------
In order to prove this theorem, we need the following lemma.
\begin{lemma}\label{lemmaw1}
Consider an open quantum system $( S, L, H)$ where $H=H_1+H_2$ and $H_2 \in \mathcal{W}_1$, and the set of non-negative self-adjoint operators $\mathcal{P}$. If there exists a $V \in \mathcal{P}$ and real constants $\tilde{\lambda}\geq0$, $\tau>0$ such that
\begin{equation}\label{lemmaw1equation}
-i[V,H_1]+\mathcal{L}(V)+\tau^2[V,z^\dag][z,V]+\frac{1}{\gamma^2\tau^2}z^\dag z+W\leq\tilde{\lambda},
\end{equation}
then
\begin{equation}
\limsup \limits_{T\rightarrow \infty} \frac{1}{T}\int_0^T\langle W(t)\rangle dt\leq\tilde{\lambda}+\frac{\delta}{\tau^2}, \forall t\geq0.
\end{equation}
\end{lemma}

\emph{Proof:}
Since $V\in \mathcal{P}$ and $H_2 \in \mathcal{W}_1$,
\begin{equation}\label{e9}
\mathcal{G}(V)=-i[V,H_1]+\mathcal{L}(V)-i[V, z^\dag]w+iw^\dag[z,V].
\end{equation}
Also,
\begin{equation}\label{e10}
\begin{split}
0&\leq(\tau[V,z^\dag]-\frac{i}{\tau}w^\dag)(\tau[V,z^\dag]-\frac{i}{\tau} w^\dag)^\dag\\
&=\tau^2[V,z^\dag][z,V]+i[V,z^\dag]w-iw^\dag[z,V]+\frac{w^\dag w}{\tau^2}.
\end{split}
\end{equation}
Substituting (\ref{e9}) into (\ref{e10}) and using the sector bound condition (\ref{sector11}), the following inequality is obtained:
\begin{equation}
\mathcal {G}(V)\leq-i[V,H_1]+\mathcal{L}(V)+\tau^2[V,z^\dag][z,V]+\frac{1}{\gamma^2\tau^2}z^\dag z+\frac{\delta}{\tau^2}.
\end{equation}
Hence,
\begin{equation}
\mathcal {G}(V)+W\leq \tilde{\lambda}+\frac{\delta}{\tau^2}.
\end{equation}
Consequently, the conclusion in the lemma follows from Lemma \ref{baselemma}.
\hfill $\Box$

%-----------------------------------
\emph{Proof of Theorem \ref{theorem1}}:
Using the Schur complement \cite{boyd}, the inequality ({\ref{riccati1}) is equivalent to
\begin{equation}\label{40}
F^\dag P+PF+4\tau^2PJE^\dag EJP+\frac{E^\dag E}{\gamma^2\tau^2}+R<0.
\end{equation}
If the Riccati inequality (\ref{40}) has a solution $P>0$ of the form (\ref{18}) and $\tau>0$, according to Lemma \ref{algebra} and Lemma \ref{quadratic_algebra}, we have
\begin{equation}
\begin{split}
&-i[V,H_1]+\mathcal{L}(V)+\tau^2[V,z^\dag][z,V]+\frac{1}{\gamma^2\tau^2}z^\dag z+W=\\
&\left[\begin{array}{l} a\\a^\# \end{array}\right]^\dag\left(\begin{array}{l}F^\dag P+PF+4\tau^2PJE^\dag EJP\\+\frac{E^\dag E}{\gamma^2\tau^2}+R\end{array}\right)\left[\begin{array}{l} a\\a^\# \end{array}\right]\\
&\quad +\text{Tr}(PJN^\dag \left[\begin{array}{cc} I&0\\0&0\end{array}\right]NJ).
\end{split}
\end{equation}
Therefore, it follows from (\ref{riccati1}) that condition (\ref{lemmaw1equation}) is satisfied with
\begin{equation}
\tilde{\lambda}=\text{Tr}(PJN^\dag \left[\begin{array}{cc} I&0\\0&0\end{array}\right]NJ)\geq0.
\end{equation}
Then, according to the relationship (\ref{w1relation}) and Lemma \ref{lemmaw1}, we have
\begin{equation}
\begin{split}
&\limsup \limits_{T\to \infty} \frac{1}{T}\int_0^T\langle W(t)\rangle dt\\
&=\limsup \limits_{T\to \infty} \frac{1}{T}\int_0^T\langle \left[\begin{array}{c c} a^\dag & a^T\end{array}\right]R\left[\begin{array}{l} a \\ a^\#\end{array}\right]\rangle dt\leq\tilde{\lambda}+\frac{\delta}{\tau^2}.
\end{split}
\end{equation}
\hfill $\Box$
%%------------------------------------------------
\subsection{Performance Analysis Using the Popov Approach}
\begin{theorem}\label{theorem11}
Consider an uncertain quantum system $(S, L, H)$, where $H=H_1+H_2$, $H_1$ is in the form of (\ref{17}),
$L$ is of the form (\ref{19}) and $H_2\in\mathcal{W}_4$. If $F=-iJM-\frac{1}{2}JN^\dag JN$ is Hurwitz, and
\begin{equation}\label{riccati}
\left[\begin{array}{cc}PF+F^\dag P+R&-2iPJE^\dag+E^\dag+\theta F^\dag E^\dag\\ 2iEJP+E+\theta EF&-\gamma I\end{array}\right]<0
\end{equation}
has a solution $P>0$ in the form of (\ref{18}) for some $\theta\geq0$, then
\begin{align}
\mathcal{J}&=\limsup \limits_{T\to \infty} \frac{1}{T}\int_0^T\langle W(t)\rangle dt\nonumber\\
&=\limsup \limits_{T\to \infty} \frac{1}{T}\int_0^T\langle \left[\begin{array}{c c} a^\dag & a^T\end{array}\right]R\left[\begin{array}{l} a \\ a^\#\end{array}\right]\rangle dt\leq\lambda\end{align}
where
\begin{equation}
\lambda=\text{Tr}(PJN^\dag \left[\begin{array}{cc} I&0\\0&0\end{array}\right]NJ)+\frac{4\theta}{\gamma}\tilde{N}^\#\Sigma JE^\dag EJ\Sigma\tilde{N}^T
.
\end{equation}
\end{theorem}
%%%
In order to prove this theorem, we need the following lemma.
\begin{lemma}\label{1lemma}
(\emph{See Theorem 1 of} \cite{Valery}) Consider a set of non-negative self-adjoint operators $\mathcal{P}$, an open quantum system $(S, L, H)$ and an observable $W$, where $H=H_1+H_2$ and $H_2\in \mathcal{W}_2$ defined in (\ref{w1}). Suppose there exists a $V\in\mathcal{P}$ and real constants $\theta\in\Theta$, $\lambda\geq0$ such that
\begin{align}
&-i[V,H_1]+\mathcal{L}(V)+\frac{1}{\gamma}(i[z,V-\theta H_1]+\theta \mathcal{L}(z)+z)^\dag\nonumber\\
&\times(i[z,V-\theta H_1]+\theta \mathcal{L}(z)+z)+\theta\beta[z,L]^\dag[z,L]+W\leq\lambda.
\end{align}
Then
\begin{equation}
\limsup \limits_{T\to \infty} \frac{1}{T}\int_0^T\langle W(t)\rangle dt\leq\lambda.
\end{equation}
Here $W(t)$ denotes the Heisenberg evolution of the operator $W$ and $\langle\cdot\rangle$ denotes quantum expectation.
\end{lemma}

%%%
\emph{Proof of Theorem \ref{theorem11}:}
Using the Schur complement, (\ref{riccati}) is equivalent to
\begin{equation}
\begin{split}
&PF+F^\dag P+\frac{1}{\gamma}(-2iPJE^\dag+E^\dag+\theta F^\dag E^\dag)\\
&\times(2iEJP+E+\theta EF)+R<0.
\end{split}
\end{equation}
According to Lemma \ref{algebra} and Lemma \ref{algebra3}, we have
\begin{align}
&-i[V,H_1]+\mathcal{L}(V)+\frac{1}{\gamma}(i[z,V-\theta H_1]+\theta \mathcal{L}(z)+z)^\dag\nonumber\\
&\times(i[z,V-\theta H_1]+\theta \mathcal{L}(z)+z)+\frac{4\theta}{\gamma}[z,L]^\dag[z,L]+W\nonumber\\
&=\left[\begin{array}{l} a\\a^\# \end{array}\right]^\dag\left(\begin{array}{l}PF+F^\dag P\\
+\frac{1}{\gamma}(-2iPJE^\dag+E^\dag+\theta F^\dag E^\dag)\\
\times(2iEJP+E+\theta EF)+R
\end{array}\right)\left[\begin{array}{l} a\\a^\# \end{array}\right]\nonumber\\
&\quad +\text{Tr}(PJN^\dag \left[\begin{array}{cc} I&0\\0&0\end{array}\right]NJ)+\frac{4\theta}{\gamma}\tilde{N}^\#\Sigma JE^\dag EJ\Sigma\tilde{N}^T.
\end{align}
From this and using the relationship (\ref{w2relation}), Lemma \ref{algebra2} and Lemma \ref{1lemma}, we obtain
\begin{equation}
\begin{split}
&\limsup \limits_{T\to \infty} \frac{1}{T}\int_0^T\langle W(t)\rangle dt\\
&=\limsup \limits_{T\to \infty} \frac{1}{T}\int_0^\infty \langle\left[\begin{array}{l} a \\ a^\#\end{array}\right]^\dag R\left[\begin{array}{l} a \\ a^\#\end{array}\right]\rangle dt\\
&\leq\lambda
\end{split}
\end{equation}
where
\begin{equation}
\lambda=\text{Tr}(PJN^\dag \left[\begin{array}{cc} I&0\\0&0\end{array}\right]NJ)+\frac{4\theta}{\gamma}\tilde{N}^\#\Sigma JE^\dag EJ\Sigma\tilde{N}^T
.
\end{equation}
\hfill $\Box$
%%---------------------------------------------------------------------------------------------------------------

%%
\section{COHERENT GUARANTEED COST CONTROLLER DESIGN}\label{controller design}
%------------------------------------------------------------------------------------
In this section, we design a coherent guaranteed cost controller for the uncertain quantum system subject to a quadratic perturbation Hamiltonian to make the control system not only stable but also to achieve an adequate level of performance. The coherent controller is realized by adding a controller Hamiltonian $H_3$. $H_3$ is assumed to be in the form
\begin{equation}\label{H3}
H_3=\frac{1}{2}\left[\begin{array}{c c} a^\dag & a^T\end{array}\right]K\left[\begin{array}{l} a \\ a^\#\end{array}\right]
\end{equation}
where $K\in\mathbb{C}^{2n\times2n}$ is a Hermitian matrix of the form
\begin{equation}
K=\left[\begin{array}{cc}K_1&K_2\\K_2^\#&K_1^\#\end{array}\right]
\end{equation}
and $K_1=K^\dag_1$, $K_2=K^T_2$.
Associated with this system is the cost function $\mathcal{J}$
\begin{equation}
%\begin{split}
%J&=\limsup \limits_{T\to \infty} \frac{1}{T}\int_0^\infty (\left[\begin{array}{l} a \\ a^\#\end{array}\right]^\dag R\left[\begin{array}{l} a \\ a^\#\end{array}\right]+\left[\begin{array}{l} a \\ a^\#\end{array}\right]^\dag \rho K^\dag K\left[\begin{array}{l} a \\ a^\#\end{array}\right])dt\\
\mathcal{J}=\limsup \limits_{T\to \infty} \frac{1}{T}\int_0^\infty \langle\left[\begin{array}{l} a \\ a^\#\end{array}\right]^\dag (R+ \rho K^2)\left[\begin{array}{l} a \\ a^\#\end{array}\right]\rangle dt
\end{equation}
where $\rho\in(0,\infty)$ is a weighting factor.
We let
\begin{equation}
W=\left[\begin{array}{l} a \\ a^\#\end{array}\right]^\dag (R+\rho K^2)\left[\begin{array}{l} a \\ a^\#\end{array}\right].
\end{equation}

%Our guaranteed cost coherent controller design method is by adding a Hamiltonian $H_3$ to the nominal system with quadratic and non-quadratic Hamiltonian perturbations, respectively,
 %and it is the main results of this paper which is shown in the following sections.
The following theorems (Theorem 3 and Theorem 4) present our main results on coherent guaranteed cost controller design for the given quantum system using a quantum small gain method and a Popov type approach, respectively.
\subsection{Coherent Controller Design Using the Small Gain Approach}
\begin{theorem}\label{theoremcontrol1}
Consider an uncertain quantum system $(S, L, H)$, where $H=H_1+H_2+H_3$, $H_1$ is in the form of (\ref{17}), $L$ is of the form (\ref{19}), $H_2\in\mathcal{W}_3$ and the controller Hamiltonian $H_3$ is in the form of (\ref{H3}). %If\\
%1)The matrix
%\begin{equation}
%F=-iJ(M+K)-\frac{1}{2}JN^\dag JN \textnormal{is Hurwits};
%\end{equation}
%and 2)
With $Q=P^{-1}$, $Y=KQ$ and $F=-iJM-\frac{1}{2}JN^\dag JN$, if there exist a matrix $Q =q*I$ ($q$ is a constant scalar and $I$ is the identity matrix), a Hermitian matrix $Y$ and a constant $\tau>0$, such that
\begin{equation}\label{lmi1}
\left[\begin{array}{cccc} A+4\tau^2JE^\dag EJ&Y& qR^{\frac{1}{2}}&qE^\dag
\\Y&-I/\rho &0&0\\qR^{\frac{1}{2}}&0&-I&0\\qE&0&0&-\gamma^2\tau^2I\end{array}\right]<0
\end{equation}
where $A=qF^\dag+Fq+iYJ-iJY$,
then the associated cost function satisfies the bound
\begin{equation}\label{optimal_controller1}
\begin{split}
&\limsup \limits_{T\to \infty} \frac{1}{T}\int_0^T\langle W(t)\rangle dt\\
&=\limsup \limits_{T\to \infty} \frac{1}{T}\int_0^\infty \langle\left[\begin{array}{l} a \\ a^\#\end{array}\right]^\dag (R+ \rho K^2)\left[\begin{array}{l} a \\ a^\#\end{array}\right]\rangle dt\\
&\leq\tilde{\lambda}+\frac{\delta}{\tau^2}
\end{split}
\end{equation}
where
\begin{equation}
\tilde{\lambda}=\text{Tr}(PJN^\dag \left[\begin{array}{cc} I&0\\0&0\end{array}\right]NJ).
\end{equation}
\end{theorem}
%%------------------------------------------------
\emph{Proof:}
Suppose the conditions of the theorem are satisfied. Using the Schur complement, (\ref{lmi1}) is equivalent to
\begin{equation}\label{eqf12}
\left[\begin{array}{ccc} A+4\tau^2JE^\dag EJ+\rho YY& qR^{\frac{1}{2}}&qE^\dag
\\qR^{\frac{1}{2}}&-I&0\\qE&0&-\gamma^2\tau^2I\end{array}\right]<0.
\end{equation}
Applying the Schur complement again, it follows that (\ref{eqf12}) is equivalent to
\begin{equation}\label{eqf111}
\left[\begin{array}{cc} A+4\tau^2JE^\dag EJ+\rho YY+q^2R&qE^\dag\\ qE&-\gamma^2\tau^2I \end{array}\right]<0
\end{equation}
and (\ref{eqf111}) is equivalent to
\begin{equation}\label{eqf11}
\begin{split}
&qF^\dag+Fq+iYJ-iJY+4\tau^2JE^\dag EJ\\
&+\rho YY+q^2(\frac{E^\dag E}{\gamma^2\tau^2}+R)<0.
\end{split}
\end{equation}
Substituting $Y=Kq=qK^\dag$ into (\ref{eqf11}), we obtain
\begin{equation}
\begin{split}
&q(F-iJK)^\dag +(F-iJK)q+4\tau^2JE^\dag EJ\\
&+q^2(\frac{E^\dag E}{\gamma^2\tau^2}+R+\rho K^2)<0.
\end{split}
\end{equation}
Since $P=Q^{-1}$, premultiplying and postmultiplying this inequality by the matrix $P$, we have
\begin{equation}\label{73}
\begin{split}
&(F-iJK)^\dag P+P(F-iJK)+4\tau^2PJE^\dag EJP\\
&+\frac{E^\dag E}{\gamma^2\tau^2}+R+\rho K^2<0.
\end{split}
\end{equation}
It follows straightforwardly from (\ref{73}) that $F-iJK$
is Hurwitz. We also know that
\begin{align}
&-i[V,H_1+H_3]+\mathcal{L}(V)+\tau^2[V,z^\dag][z,V]+\frac{1}{\gamma^2\tau^2}z^\dag z+W\nonumber\\
&=\left[\begin{array}{l} a\\a^\# \end{array}\right]^\dag\left(\begin{array}{l}(F-iJK)^\dag P+P(F-iJK)\nonumber\\
+4\tau^2PJE^\dag EJP+E^\dag E/(\gamma^2\tau^2)\\+R+\rho K^2
\end{array}\right)\left[\begin{array}{l} a\\a^\# \end{array}\right]\nonumber\\
&\quad +\text{Tr}(PJN^\dag \left[\begin{array}{cc} I&0\\0&0\end{array}\right]NJ).
\end{align}
According to the relationship (\ref{w1relation}) and Lemma \ref{lemmaw1}, we have
\begin{equation}
\begin{split}
&\limsup \limits_{T\to \infty} \frac{1}{T}\int_0^T\langle W(t)\rangle dt\\
&=\limsup \limits_{T\to \infty} \frac{1}{T}\int_0^\infty \langle\left[\begin{array}{l} a \\ a^\#\end{array}\right]^\dag (R+ \rho K^2)\left[\begin{array}{l} a \\ a^\#\end{array}\right]\rangle dt\\
&\leq\tilde{\lambda}+\frac{\delta}{\tau^2}
\end{split}
\end{equation}
where
\begin{equation}
\tilde{\lambda}=\text{Tr}(PJN^\dag \left[\begin{array}{cc} I&0\\0&0\end{array}\right]NJ).
\end{equation}
\hfill $\Box$
\begin{remark}\label{remark1}
In order to design a coherent controller which minimizes the cost bound (\ref{optimal_controller1}) in Theorem 3, we need to formulate an inequality
\begin{equation}\label{quadraticxi}
\text{Tr}(PJN^\dag \left[\begin{array}{cc} I&0\\0&0\end{array}\right]NJ)+\frac{\delta}{\tau^2}\leq\xi.
\end{equation}
We know that $P=Q^{-1}=q^{-1}I$ and apply the Schur complement to inequality (\ref{quadraticxi}), so that we have
\begin{equation}\label{qudraticxi1}
\left[\begin{array}{cc} -\xi+\frac{\delta}{\tau^2}&B^\frac{1}{2}\\ B^\frac{1}{2}&-q\end{array}\right]\leq0
\end{equation}
where $B=\text{Tr}(JN^\dag \left[\begin{array}{cc} I&0\\0&0\end{array}\right]NJ).$
Applying the Schur complement again, it is clear that (\ref{qudraticxi1}) is equivalent to
\begin{equation}\label{qudraticxi2}
\left[\begin{array}{ccc} -\xi&\delta^\frac{1}{2}&B^\frac{1}{2}\\ \delta^\frac{1}{2}&-\tau^2&0\\B^\frac{1}{2}&0&-q\end{array}\right]\leq0.
\end{equation}
Hence, we minimize $\xi$ subject to  (\ref{qudraticxi2}) and (\ref{lmi1}) in Theorem \ref{theoremcontrol1}. This is a standard LMI problem.
%Essentially, we try to minimise the cost bound
%$\text{Tr}(PJN^\dag \left[\begin{array}{cc} I&0\\0&0\end{array}\right]NJ)$ subject to (\ref{lmi}) in the above theorem.
\end{remark}
\subsection{Coherent Controller Design Using the Popov Approach}
\begin{theorem}\label{theoremcontrol2}
Consider an uncertain quantum system $(S, L, H)$, where $H=H_1+H_2+H_3$, $H_1$ is in the form of (\ref{17}), $L$ is of the form (\ref{19}), $H_2\in\mathcal{W}_4$, the controller Hamiltonian $H_3$ is in the form of (\ref{H3}). With $Q=P^{-1}$, $Y=KQ$ and $F=-iJM-\frac{1}{2}JN^\dag JN$, if there exist a matrix $Q =q*I$ ($q$ is a constant scalar and $I$ is the identity matrix), a Hermitian matrix $Y$ and a constant $\theta>0$, such that
\begin{equation}\label{lmi}
\left[\begin{array}{cccc}A&B^\dag&Y&qR^\frac{1}{2}\\B&-\gamma I&0&0\\Y&0&-I/\rho&0\\qR^\frac{1}{2}&0&0&-I\end{array}\right]<0
\end{equation}
where $A=Fq+qF^\dag-iJY+iYJ$ and $B=2iEJ +Eq+\theta EFq-i\theta EJY$,
then the associated cost function satisfies the bound
\begin{equation}\label{optimal_controller}
\begin{split}
&\limsup \limits_{T\to \infty} \frac{1}{T}\int_0^T\langle W(t)\rangle dt\\
&=\limsup \limits_{T\to \infty} \frac{1}{T}\int_0^\infty \langle\left[\begin{array}{l} a \\ a^\#\end{array}\right]^\dag (R+ \rho K^2)\left[\begin{array}{l} a \\ a^\#\end{array}\right]\rangle dt\leq\lambda
\end{split}
\end{equation}
where
\begin{equation}\label{thebound}
\lambda=\text{Tr}(PJN^\dag \left[\begin{array}{cc} I&0\\0&0\end{array}\right]NJ)+\frac{4\theta}{\gamma}\tilde{N}^\#\Sigma JE^\dag EJ\Sigma\tilde{N}^T.
\end{equation}
\end{theorem}
%%------------------------------------------------
\emph{Proof:}
Suppose the conditions of the theorem are satisfied. Using the Schur complement, (\ref{lmi}) is equivalent to
\begin{equation}\label{first}
\left[\begin{array}{ccc}A+\frac{1}{\gamma}B^\dag B&Y&qR^\frac{1}{2}\\Y&-I/\rho&0\\qR^\frac{1}{2}&0&-I\end{array}\right]<0.
\end{equation}
We then apply the Schur complement to inequality (\ref{first}) and obtain
\begin{equation}\label{second}
\left[\begin{array}{cc}A+\frac{1}{\gamma}B^\dag B+\rho YY&qR^\frac{1}{2}\\qR^\frac{1}{2}&-I\end{array}\right]<0.
\end{equation}
Also, (\ref{second}) is equivalent to
\begin{equation}\label{eqf1}
\begin{split}
&Fq+qF^\dag-iJY+iYJ+\\
&\frac{1}{\gamma}(-2iJE^\dag+qE^\dag+\theta qF^\dag E^\dag+i\theta YJE^\dag)\\
&\times(2iEJ +Eq+\theta EFq-i\theta EJY)\\
&+q^2R+\rho YY<0.
\end{split}
\end{equation}
Substituting $Y=Kq=qK^\dag$ into (\ref{eqf1}), we obtain
\begin{equation}
\begin{split}
&(F-iJK)q+q(F-iJK)^\dag\\
&+\frac{1}{\gamma}(-2iJE^\dag+qE^\dag+\theta q(F-iJK)^\dag E^\dag)\\
&\times(2iEJ+Eq+\theta E(F-iJK)q)\\
&+q^2(R+\rho K^2)<0.
\end{split}
\end{equation}
Since $P=Q^{-1}$, premultiplying and postmultiplying this inequality by the matrix $P$, we have
\begin{equation}\label{87}
\begin{split}
&P(F-iJK)+(F-iJK)^\dag P\\
&+\frac{1}{\gamma}(-2iPJE^\dag+E^\dag+\theta (F-iJK)^\dag E^\dag)\\
&\times(2iEJP+E+\theta E(F-iJK))+R+\rho K^2<0.
\end{split}
\end{equation}
%We know $E^\dag E\geq0, R>0$ and $K^2>0$. Hence\\
%\begin{equation}
%\begin{split}
%4\tau^2PJE^\dag EJP+\frac{E^\dag E}{\gamma^2\tau^2}+R+\rho K^2>0,\\
%(F-iJK)^\dag P+P(F-iJK)<0.
%\end{split}
%\end{equation}
%Therefore, we have the following fact
%\begin{equation}
%\tilde{F}=-iJ(M+K)-\frac{1}{2}JN^\dag JN \textnormal{ is Hurwitz}.
%\end{equation}
It follows straightforwardly from (\ref{87}) that $F-iJK$ is Hurwitz.
We also know that
\begin{align}
&-i[V,H_1+H_3]+\mathcal{L}(V)\nonumber\\
&+\frac{1}{\gamma}(i[z,V-\theta (H_1+H_3)]+\theta \mathcal{L}(z)+z)^\dag\nonumber\\
&\times(i[z,V-\theta (H_1+H_3)]+\theta \mathcal{L}(z)+z)\nonumber\\
&+\frac{4\theta}{\gamma}[z,L]^\dag[z,L]+W\nonumber\\
&=\left[\begin{array}{l} a\\a^\# \end{array}\right]^\dag\tilde{M}\left[\begin{array}{l} a\\a^\# \end{array}\right]\nonumber\\
&\quad +\text{Tr}(PJN^\dag \left[\begin{array}{cc} I&0\\0&0\end{array}\right]NJ)+\frac{4\theta}{\gamma}\tilde{N}^\#\Sigma JE^\dag EJ\Sigma\tilde{N}^T
\end{align}
where
\begin{equation}
\begin{split}
\tilde{M}=
&P(F-iJK)+(F-iJK)^\dag P\\
&+\frac{1}{\gamma}(-2iPJE^\dag+E^\dag+\theta (F-iJK)^\dag E^\dag)\\
&\times(2iEJP+E+\theta E(F-iJK))\\
&+R+\rho K^2.
\end{split}
\end{equation}
According to the relationship (\ref{w2relation}) and Lemma \ref{1lemma}, we have
\begin{equation}
\begin{split}
&\limsup \limits_{T\to \infty} \frac{1}{T}\int_0^T\langle W(t)\rangle dt\\
&=\limsup \limits_{T\to \infty} \frac{1}{T}\int_0^\infty \langle\left[\begin{array}{l} a \\ a^\#\end{array}\right]^\dag (R+ \rho K^2)\left[\begin{array}{l} a \\ a^\#\end{array}\right]\rangle dt\\
&\leq\lambda
\end{split}
\end{equation}
where
\begin{equation}
\lambda=\text{Tr}(PJN^\dag \left[\begin{array}{cc} I&0\\0&0\end{array}\right]NJ)+\frac{4\theta}{\gamma}\tilde{N}^\#\Sigma JE^\dag EJ\Sigma\tilde{N}^T
.
\end{equation}
\hfill $\Box$

\begin{remark}
For each fixed value of $\theta$, the problem is an LMI problem. Then, we can iterate on $\theta \in [0,\infty)$ and choose the value which minimizes the cost bound (\ref{thebound}) in Theorem \ref{theoremcontrol2}.
\end{remark}
%%
%%----------------------------------------------------------------------------------------------------------------

%
\section{ILLUSTRATIVE EXAMPLE}\label{illustrative example}
%---------------------------------------------------------------------------------------------------
In order to illustrate our methods and compare their performance, we use the same quantum system considered in \cite{Chengdi2} as an example. The system corresponds to a degenerate parametric amplifier and its $(S,L,H)$ description has the following form
\begin{equation}\label{SLH}
H=\frac{1}{2}i((a^\dag)^2-a^2),
S=I,
L=\sqrt{\kappa}a.
\end{equation}
We let the perturbation Hamiltonian be
\begin{equation}
H_2=\frac{1}{2}\left[\begin{array}{c c} a^\dag & a^T\end{array}\right]\left[\begin{array}{c c}1&0.5i\\-0.5i&1\end{array}\right]\left[\begin{array}{l} a \\ a^\#\end{array}\right]
\end{equation}
and the nominal Hamiltonian be
\begin{equation}
H_1=\frac{1}{2}\left[\begin{array}{c c} a^\dag & a^T\end{array}\right]\left[\begin{array}{c c}-1&0.5i\\-0.5i&-1\end{array}\right]\left[\begin{array}{l} a \\ a^\#\end{array}\right]
\end{equation}
so that $H_1+H_2=H$.
The corresponding parameters considered in Theorem 1, Theorem 2, Theorem \ref{theoremcontrol1} and Theorem \ref{theoremcontrol2} are as follows:
\begin{equation}
\begin{split}
&M=\left[\begin{array}{cc} -1&0.5i \\ -0.5i&-1\end{array}\right],
N=\left[\begin{array}{cc} \sqrt{\kappa}&0\\ 0&\sqrt{\kappa}\end{array}\right],\\
&F=\left[\begin{array}{cc} -\frac{\kappa}{2}+i&0.5\\ 0.5&-\frac{\kappa}{2}-i\end{array}\right],
E=I
\end{split}
\end{equation}
and
\begin{equation}
\Delta=\left[\begin{array}{c c}1&0.5i\\-0.5i&1\end{array}\right].
\end{equation}
%\begin{equation}
%\Delta\Delta=\left[\begin{array}{cc} 0&\frac{1}{2}i \\ -\frac{1}{2}i&0\end{array}\right]\left[\begin{array}{cc} %0&\frac{1}{2}i \\ -\frac{1}{2}i&0\end{array}\right]=\left[\begin{array}{cc} \frac{1}{4} &0\\ %0&\frac{1}{4}\end{array}\right].
%\end{equation}

To illustrate Theorem 1 and Theorem 3, we consider $H_2\in \mathcal{W}_3$. Hence, $\gamma=1$ is chosen to satisfy (\ref{delta_bound1}). The performance using the small gain approach for the uncertain quantum system is shown in Figure 1. In Figure 1, the dashed line represents the cost bound for the linear quantum system considered in Theorem 1 as a function of the parameter $\kappa$. The solid line shows the system performance with the coherent controller designed in Theorem 3. Compared to the performance without a controller, the coherent controller can guarantee that the system is stable for a larger range of the damping parameter $\kappa$ and gives the system improved performance.

Now we illustrate one approach to realizing the desired controller. For instance, when $\kappa=4.5$, by using the controller design method in Theorem 3, we have the desired controller Hamiltonian as 
\begin{equation}\label{last1}
H_3=\frac{1}{2}\left[\begin{array}{c c} a^\dag & a^T\end{array}\right]\left[\begin{array}{cc} 0&-0.5i \\ 0.5i&0\end{array}\right]\left[\begin{array}{l} a \\ a^\#\end{array}\right].
\end{equation}
This controller Hamiltonian can be realized by connecting the degenerate parametric amplifier with a static squeezer as shown in Figure 2. This static squeezer is a static Bogoliubov component which corresponds to the Bogoliubov transformation \cite{Gough2010squeezing}, \cite{Vuglar2012}. Also, we have the following definition:
\begin{definition} (see \cite{Gough2010squeezing}, \cite{Vuglar2012}) 
A static Bogoliubov component is a component that implements the Bogoliubov transformation:
$\left[\begin{array}{c} dy(t)\\dy^\#(t)\end{array}\right]
=B\left[\begin{array}{c} du(t)\\du^\#(t)\end{array}\right],$
where 
$B=\left[\begin{array}{c c} B_1&B_2\\B^\#_2&B^\#_1 \end{array}\right],
B^\dag JB=J.$
\end{definition}
To realize $H_3$ in (\ref{last1}), we let the matrix $B=\left[\begin{array}{r r}\frac{5}{4}&-\frac{3}{4}\\-\frac{3}{4}&\frac{5}{4} \end{array}\right]$ which satisfies the Bogoliubov condition $B^\dag J B=J$, and $\tilde{\kappa}=\frac{1}{3}$. Therefore, the overall Hamiltonian of the closed loop system is $H=H_1+H_2+H_3$, which achieves the controller design goal. Detailed procedure regarding how to get the matrix $B$ can be found in appendix. 
\begin{figure}[htb]
       \centering
        \includegraphics[width=0.5  \textwidth]{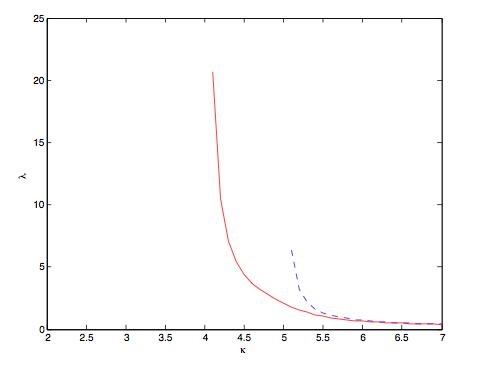}
        \caption{Guaranteed cost bounds for an uncertain quantum system with a controller (solid line) and without a controller (dashed line) using the small gain approach}
        \label{fig1}
\end{figure}
%------------------------------------------------
\begin{figure}[htb]
       \centering
        \includegraphics[width=0.5  \textwidth]{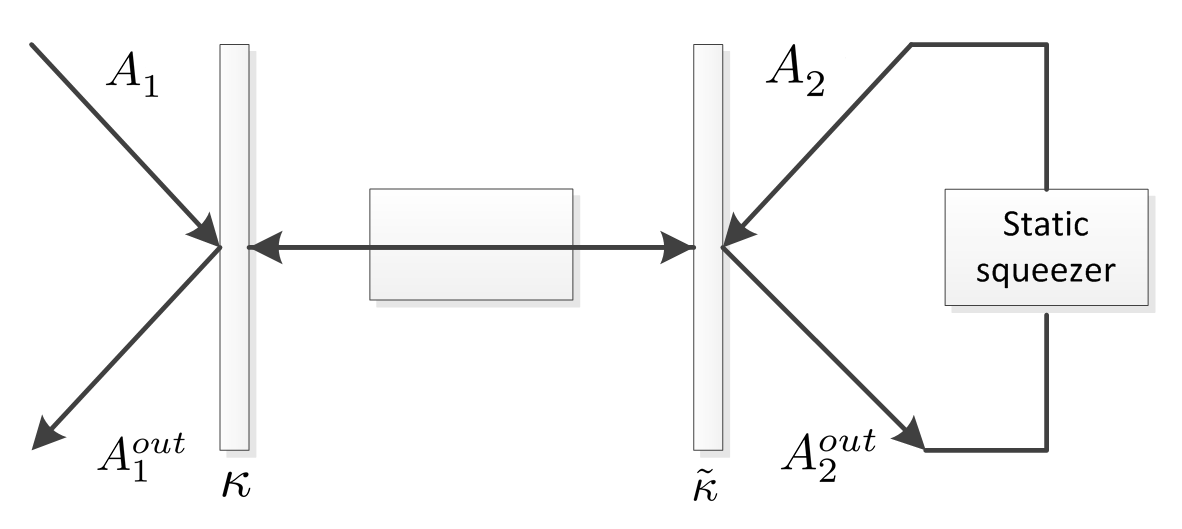}
        \caption{Degenerate parametric amplifier coupled to a static squeezer.}
\end{figure}

To illustrate Theorem 2 and Theorem 4, we consider $H_2\in \mathcal{W}_4$. Hence, $\gamma=2$ is chosen to satisfy (\ref{delta_bound}).  The results using the Popov approach are shown in Figure 3 and Figure 4. Figure 3 demonstrates how to choose the value of $\theta$. We consider the same example as above with $\kappa=3.8$ and iterate on $\theta\in[0,1]$. Figure 3 shows the cost bound for this quantum system obtained in Theorem 4 as a function of the parameter $\theta$. It is clear that the minimal cost bound is achieved when $\theta=0.1$. Therefore, we choose $\theta=0.1$ for $\kappa=3.8$ and use a similar method to choose $\theta$ for other values of $\kappa$.
\begin{figure}[htb]%kappa is 3.8
       \centering
        \includegraphics[width=0.5  \textwidth]{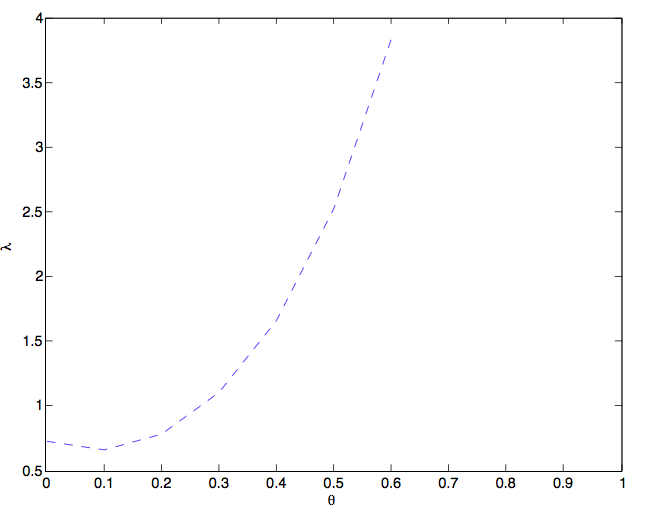}
        \caption{Performance versus $\theta$}
        \label{fig1}
\end{figure}

In Figure 4, the dashed line shows the performance for the given system considered in Theorem 2 and the solid line describes the cost bound for the linear quantum system with the coherent controller considered in Theorem 4. As can be seen in Figure 4, the system with a controller has better performance than the case without a controller.

\begin{figure}[htb]
       \centering
        \includegraphics[width=0.5  \textwidth]{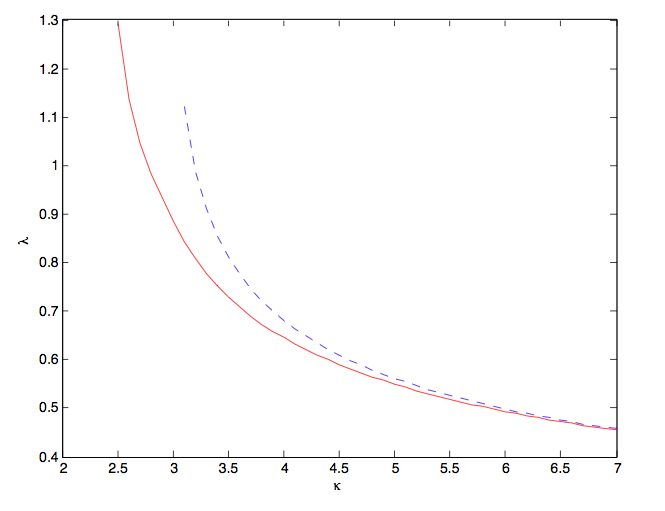}
        \caption{Guaranteed cost bounds for the uncertain quantum system with a controller (solid line) and without a controller (dashed line) using the Popov approach}
        \label{fig1}
\end{figure}
%---------------------------------------------------
Also, we can observe that the method in Theorem 3 can only make the quantum system stable for $\kappa>4$ in the example. Therefore, compared with the results in Figure 1, the Popov method obtains a lower cost bound and a larger range of robust stability as shown in Figure 4. This is as expected, since the Popov approach allows for a more general class of Lyapunov functions than the small gain approach.
%%%%%%%%%%%%%%%%

%To illustrate Remark 2, we consider the same example as above with $\kappa=3.8$ and iterate on $\theta\in[0,1]$. Figure 3 shows the cost bound for this quantum system obtained in Theorem 4 as a function of parameter $\theta$. It is clear that the minimal cost bound is achieved when $\theta=0.1$.

%----------------------------------------------------------------------------------------------------------------------

\section{CONCLUSION}\label{conclusion}

In this paper, the small gain method and the Popov approach, respectively, are used to analyze the performance of an uncertain linear quantum system subject to a quadratic perturbation in the system Hamiltonian. Then, we add a coherent controller to make the given system not only stable but also to achieve improved performance.
By an illustrative example, we have also shown that the Popov method adds to a considerable improvement over the small gain method in terms of system performance. Future work will include the extension of these approaches to nonlinear uncertain quantum systems \cite{IanPortland}.

%--------------------------------------------------------------------------------------------

\appendix
The detailed procedure regarding how to realize the desired controller is shown below. We consider a degenerate parametric amplifier (DPA) as an example. Based on the $(S,L,H)$ description in (\ref{SLH}), we can calculate the following quantum stochastic differential equations \cite{matt2008} describing the DPA:
\begin{equation}
\begin{split}
%da(t)&=-\frac{\kappa}{2}a dt+\frac{a^\ast}{2} dt-\sqrt{\kappa}dA_1(t);\\
\left[\begin{array}{c} da(t)\\da^\#(t)\end{array}\right]=&\left[\begin{array}{c c}-\frac{\kappa}{2}&1\\1&-\frac{\kappa}{2}\end{array}\right]\left[\begin{array}{c} a(t)\\a^\#(t)\end{array}\right]dt\\
&-\left[\begin{array}{c c} \sqrt{\kappa}&0\\0&\sqrt{\kappa}\end{array}\right]\left[\begin{array}{c} dA_1(t)\\dA^\#_1(t)\end{array}\right];\\
dA^\text{out}_1(t)=&\sqrt{\kappa}a(t)dt+dA_1(t).\nonumber
\end{split}
\end{equation}
We have known that when $\kappa=4.5$ and using the controller design method in Theorem 3, we have the desired controller Hamiltonian as in (\ref{last1}).

Next, we show how to realize this controller Hamiltonian by connecting this degenerate parametric amplifier with a static squeezer as shown in Fig. 2. The corresponding quantum stochastic differential equations for this DPA is as follows:
\begin{equation}\label{3}
\begin{split}
%da(t)&=-\frac{\gamma}{2}a dt+\frac{a^\ast}{2} dt-\sqrt{\kappa}dA_1(t)-%%\sqrt{\tilde{\kappa}}dA_2(t);\\
\left[\begin{array}{c} da(t)\\da^\#(t)\end{array}\right]&=\left[\begin{array}{c c}-\frac{\kappa+\tilde{\kappa}}{2}&1\\1&-\frac{\kappa+\tilde{\kappa}}{2}\end{array}\right]\left[\begin{array}{c} a(t)\\a^\#(t)\end{array}\right]dt\\
&-\left[\begin{array}{c c} \sqrt{\kappa}&0\\0&\sqrt{\kappa}\end{array}\right]\left[\begin{array}{c} dA_1(t)\\dA^\#_1(t)\end{array}\right]\\
&\quad-\left[\begin{array}{c c} \sqrt{\tilde{\kappa}}&0\\0&\sqrt{\tilde{\kappa}}\end{array}\right]\left[\begin{array}{c} dA_2(t)\\dA^\#_2(t)\end{array}\right];\\
dA^\text{out}_1(t)&=\sqrt{\kappa}a(t)dt+dA_1(t);\\
dA^\text{out}_2(t)&=\sqrt{\tilde{\kappa}}a(t)dt+dA_2(t).
\end{split}
\end{equation}
We have known that this static squeezer is a static Bogoliubov component which satisfies Definition 1. According to the Definition 1, we have the following relation:
\begin{equation}\label{5}
\left[\begin{array}{c} dA_2(t)\\dA^\#_2(t)\end{array}\right]=B\left[\begin{array}{c} dA^\text{out}_2(t)\\dA_2^{\text{out}\#}(t)\end{array}\right].
\end{equation}

According to (\ref{3}), we have 
\begin{equation}\label{4}
\begin{split}
\left[\begin{array}{c} dA^\text{out}_2(t)\\dA^{\text{out}\#}_2(t)\end{array}\right]=&\left[\begin{array}{c c} \sqrt{\tilde{\kappa}}&0\\0&\sqrt{\tilde{\kappa}}\end{array}\right]\left[\begin{array}{c} a(t)\\a^\#(t)\end{array}\right]dt\\
&+\left[\begin{array}{c} dA_2(t)\\dA_2^\#(t)\end{array}\right].
\end{split}
\end{equation}

Substituting (\ref{5}) into (\ref{4}), we obtain that 
\begin{equation}
\begin{split}
B^{-1}\left[\begin{array}{c} dA_2(t)\\dA^\#_2(t)\end{array}\right]=&\left[\begin{array}{c c} \sqrt{\tilde{\kappa}}&0\\0&\sqrt{\tilde{\kappa}}\end{array}\right]\left[\begin{array}{c} a(t)\\a^\#(t)\end{array}\right]dt\\
&+\left[\begin{array}{c} dA_2(t)\\dA_2^\#(t)\end{array}\right].\nonumber
\end{split}
\end{equation}
Hence,
\begin{equation}\label{6}
(B^{-1}-I)\left[\begin{array}{c} dA_2(t)\\dA^\#_2(t)\end{array}\right]=\left[\begin{array}{c c} \sqrt{\tilde{\kappa}}&0\\0&\sqrt{\tilde{\kappa}}\end{array}\right]\left[\begin{array}{c} a(t)\\a^\#(t)\end{array}\right]dt.
\end{equation}
We now assume inverse of $B^{-1}-I$ exists. It follows (\ref{6}) that we can write
\begin{equation}\label{6.1}
\left[\begin{array}{c} dA_2(t)\\dA^\#_2(t)\end{array}\right]=(B^{-1}-I)^{-1}\left[\begin{array}{c c} \sqrt{\tilde{\kappa}}&0\\0&\sqrt{\tilde{\kappa}}\end{array}\right]\left[\begin{array}{c} a(t)\\a^\#(t)\end{array}\right]dt.
\end{equation}
Substituting (\ref{6.1}) into first equation in (\ref{3}), we get
\begin{equation}\label{7}
\begin{split}
\left[\begin{array}{c} da(t)\\da^\#(t)\end{array}\right]=&\left[\begin{array}{c c}-\frac{\kappa+\tilde{\kappa}}{2}&1\\1&-\frac{\kappa+\tilde{\kappa}}{2}\end{array}\right]\left[\begin{array}{c} a(t)\\a^\#(t)\end{array}\right]dt\\
&-\left[\begin{array}{c c} \sqrt{\kappa}&0\\0&\sqrt{\kappa}\end{array}\right]\left[\begin{array}{c} dA_1(t)\\dA^\#_1(t)\end{array}\right]\\
&-\left[\begin{array}{c c} \sqrt{\tilde{\kappa}}&0\\0&\sqrt{\tilde{\kappa}}\end{array}\right](B^{-1}-I)^{-1}\\
&\times\left[\begin{array}{c c} \sqrt{\tilde{\kappa}}&0\\0&\sqrt{\tilde{\kappa}}\end{array}\right]\left[\begin{array}{c} a(t)\\a^\#(t)\end{array}\right]dt\\
=&\left[\begin{array}{c c}-\frac{\kappa+\tilde{\kappa}}{2}&1\\1&-\frac{\kappa+\tilde{\kappa}}{2}\end{array}\right]\left[\begin{array}{c} a(t)\\a^\#(t)\end{array}\right]dt\\
&-\left[\begin{array}{c c} \sqrt{\kappa}&0\\0&\sqrt{\kappa}\end{array}\right]\left[\begin{array}{c} dA_1(t)\\dA^\#_1(t)\end{array}\right]\\
&-(B^{-1}-I)^{-1}\left[\begin{array}{c c} \tilde{\kappa}&0\\0&\tilde{\kappa}\end{array}\right]\left[\begin{array}{c} a(t)\\a^\#(t)\end{array}\right]dt.
\end{split}
\end{equation}

According to Definition 1, the $B$ matrix satisfies the following relation:
\begin{equation}
\begin{split}
JB^\dag JB&=\left[\begin{array}{c c} I&0\\0&-I\end{array}\right]\left[\begin{array}{c c} B^\dag_1&B_2^T\\B_2^\dag&B_1^T\end{array}\right]\left[\begin{array}{c c} I&0\\0&-I\end{array}\right]\left[\begin{array}{c c} B_1&B_2\\B_2^\#&B_1^\#\end{array}\right]\\
&=\left[\begin{array}{c c} B^\dag_1&B_2^T\\-B_2^\dag&-B_1^T\end{array}\right]\left[\begin{array}{c c} B_1&B_2\\-B_2^\#&-B_1^\#\end{array}\right]\\
&=\left[\begin{array}{c c} B^\dag_1B_1-B_2^TB_2^\#& B_1^\dag B_2-B_2^TB_1^\#\\-B_2^\dag B_1+B_1^TB_2^\#&-B_2^\dag B_2+B_1^TB_1^\#\end{array}\right]=I.\nonumber
\end{split}
\end{equation}
In our case that $B$ is a $2\times 2$ matrix, we have 
\begin{equation}
B_1^\dag B_2-B_2^TB_1^\#=0; -B_2^\dag B_1+B_1^TB_2^\#=0.\nonumber
\end{equation}
Moreover, we need to have the following relation:
\small
\begin{equation}
B_1^\dag B_1-B_2^TB_2^\#=-B_2^\dag B_2+B_1^TB_1^\#=(B_{1x}^2+B_{1y}^2)-(B_{2x}^2+B_{2y}^2)=I.\nonumber
\end{equation}
\normalsize
where $ B_1=B_{1x}+iB_{1y}, B_2=B_{2x}+iB_{2y}$.
Thus, we can assume that 
\begin{equation}
B_{1x}^2+B_{1y}^2=\text{cosh}(r)^2;
B_{2x}^2+B_{2y}^2=\text{sinh}(r)^2.\nonumber
\end{equation}
Hence, we may write the matrix $B$ in the following form:
\footnotesize
\begin{equation}
B=\left[\begin{array}{c c} \text{cosh}(r)\text{cos}(\alpha)+i\text{cosh}(r)\text{sin}(\alpha)&
\text{sinh}(r)\text{cos}(\beta)+i\text{sinh}(r)\text{sin}(\beta)\\\text{sinh}(r)\text{cos}(\beta)-i\text{sinh}(r)\text{sin}(\beta)&\text{cosh}(r)\text{cos}(\alpha)-i\text{cosh}(r)\text{sin}(\alpha)\end{array}\right].\nonumber
\end{equation}
\normalsize
Since $B,I$ are  $2\times 2$ matrices, we have 
\begin{equation}
(B^{-1}-I)^{-1}=\frac{1}{2-2B_{1x}}\left[\begin{array}{c c} B_1-1&B_2\\B^\#_2&B_1^\#-1\end{array}\right]. \nonumber
\end{equation}

Therefore, the last term on the right side of equation (\ref{7}) can be expressed as:
\scriptsize
\begin{equation}\label{8}
\begin{split}
&\tilde{\kappa}(B^{-1}-I)^{-1}=\frac{\tilde{\kappa}}{2-2\text{cosh}(r)\text{cos}(\alpha)}\\
&\times(\left[\begin{array}{c c} \text{cosh}(r)\text{cos}(\alpha)-1&
0\\0&\text{cosh}(r)\text{cos}(\alpha)-1\end{array}\right]\\
&+\left[\begin{array}{c c} i\text{cosh}(r)\text{sin}(\alpha)&
\text{sinh}(r)\text{cos}(\beta)+i\text{sinh}(r)\text{sin}(\beta)\\\text{sinh}(r)\text{cos}(\beta)-i\text{sinh}(r)\text{sin}(\beta)&-i\text{cosh}(r)\text{sin}(\alpha)\end{array}\right])\\
=&\left[\begin{array}{c c} -\frac{\tilde{\kappa}}{2}&
0\\0&-\frac{\tilde{\kappa}}{2}\end{array}\right]+\frac{\tilde{\kappa}}{2-2\text{cosh}(r)\text{cos}(\alpha)}\\
&\times\left[\begin{array}{c c} i\text{cosh}(r)\text{sin}(\alpha)&
\text{sinh}(r)\text{cos}(\beta)+i\text{sinh}(r)\text{sin}(\beta)\\\text{sinh}(r)\text{cos}(\beta)-i\text{sinh}(r)\text{sin}(\beta)&-i\text{cosh}(r)\text{sin}(\alpha)\end{array}\right].
\end{split}
\end{equation}
\normalsize
Substituting (\ref{8}) into (\ref{7}), we have
\footnotesize
\begin{equation}\label{9}
\begin{split}
&\left[\begin{array}{c} da(t)\\da^\#(t)\end{array}\right]=\left[\begin{array}{c c}-\frac{\kappa}{2}&1\\1&-\frac{\kappa}{2}\end{array}\right]\left[\begin{array}{c} a(t)\\a^\#(t)\end{array}\right]dt\\
&-\left[\begin{array}{c c} \sqrt{\kappa}&0\\0&\sqrt{\kappa}\end{array}\right]\left[\begin{array}{c} dA_1(t)\\dA^\#_1(t)\end{array}\right]\\
&-\frac{\tilde{\kappa}}{2-2\text{cosh}(r)\text{cos}(\alpha)}\times\\
&\left[\begin{array}{c c} i\text{cosh}(r)\text{sin}(\alpha)&
\text{sinh}(r)\text{cos}(\beta)+i\text{sinh}(r)\text{sin}(\beta)\\\text{sinh}(r)\text{cos}(\beta)-i\text{sinh}(r)\text{sin}(\beta)&-i\text{cosh}(r)\text{sin}(\alpha)\end{array}\right]\\
&\times\left[\begin{array}{c} a(t)\\a^\#(t)\end{array}\right]dt.
\end{split}
\end{equation}
\normalsize
Therefore, the closed loop system with DPA and a static squeezer has the dynamical equation (\ref{9}). The difference between this closed loop system (\ref{9}) and the original uncertain quantum system (\ref{3}) is the addition of the last term on right side of (\ref{9}) which corresponds to the controller Hamiltonian $H_3$. To realize our desired controller Hamiltonian
$H_3$ as in (\ref{last1}), the following relation is required:
\footnotesize
\begin{equation}\label{10}
\begin{split}
&-\frac{\tilde{\kappa}}{2-2\text{cosh}(r)\text{cos}(\alpha)}\times\\&\left[\begin{array}{c c} i\text{cosh}(r)\text{sin}(\alpha)&
\text{sinh}(r)\text{cos}(\beta)+i\text{sinh}(r)\text{sin}(\beta)\\\text{sinh}(r)\text{cos}(\beta)-i\text{sinh}(r)\text{sin}(\beta)&-i\text{cosh}(r)\text{sin}(\alpha)\end{array}\right]\\
&=-iJK=\left[\begin{array}{c c}0&-0.5\\-0.5&0\end{array}\right].
\end{split}
\end{equation}
\normalsize
When $\alpha=0, \beta=0,\text{sinh}(r)=-\frac{3}{4}, \text{cosh}(r)=\frac{5}{4}, \tilde{\kappa}=\frac{1}{3}$, the relationship (\ref{10}) holds.
In this case, we may have $B$ matrix as $B=\left[\begin{array}{r r}\frac{5}{4}&-\frac{3}{4}\\-\frac{3}{4}&\frac{5}{4} \end{array}\right]$ which satisfies the Bogoliubov condition
$B^\dag J B=J$.

% that's all folks
\end{document}